# Micro-crystalline inclusions analysis by PIXE and RBS

D. Strivay[a,b], C. Ramboz[b], J.-P. Gallien[c], D. Grambole[d], T. Sauvage[e] and K. Kouzmanov[b,f]

[a]Institut de Physique Nucléaire, Atomique et Spectroscopie, Université de Liège, Belgium

[b]Institut des Sciences de la Terre d'Orléans, CNRS-Orléans, France

[c]Laboratoire Pierre Süe, CEA-Saclay, France

[d]Ionenstrahlzentrum, Forschungszentrum Rossendorf, Germany

[e]Centre d'Etudes et de Recherches par Irradiation, CNRS-Orléans, France

[f]Sciences de la Terre, Université de Genève, France

## Abstract

A characteristic feature of the nuclear microprobe using a 3 MeV proton beam is the long range of particles (around 70 μm in light matrices). The PIXE method, with EDS analysis and using the multilayer approach for treating the X-ray spectrum allows the chemistry of an intra-crystalline inclusion to be measured, provided the inclusion roof and thickness at the impact point of the beam ($Z$ and $e$, respectively) are known (the depth of the inclusion floor is $Z + e$). The parameter $Z$ of an inclusion in a mineral can be measured with a precision of around 1 μm using a motorized microscope. However, this value may significantly depart from $Z$ if the analyzed inclusion has a complex shape. The parameter $e$ can hardly be measured optically. By using combined RBS and PIXE measurements, it is possible to obtain the geometrical information needed for quantitative elemental analysis. This paper will present measurements on synthetic samples to investigate the advantages of the technique, and also on natural solid and fluid inclusions in quartz. The influence of the geometrical parameters will be discussed with regard to the concentration determination by PIXE. In particular, accuracy of monazite micro-inclusion dating by coupled PIXE–RBS will be presented.

**Keywords:** Micro-inclusion; RBS; PIXE; Microprobe; Monazite

## 1. Introduction

Intra-crystalline solid or fluid micro-inclusions (size between 1 and 30 μm) are unique witnesses of pressure–temperature–time conditions of rock formation. They are helpful to analyze the chemical history of a rock. Major and trace element analysis is needed to characterize them. A specific feature of the PIXE (particle Induced X-ray emission) method is its ability to analyse intra-crystalline micro-inclusions in a non-destructive way [1], [2], [3] and [4]. This is because the range of 3 MeV protons is of the order of 70 μm in silicates. However, to be fully quantitative, the PIXE method requires the exact knowledge of the geometrical features of the layers successively passed through by the beam, both outside and within the target. When analysing intra-crystalline micro-inclusions, knowledge of the geometry and the chemical composition of the layers passed through by the protons and photons within the matrix become essential for controlling the accuracy of the analysis of the irradiated phases separately. The quantitative analysis of intra-crystalline inclusions in geological targets is commonly performed using the multilayer yield model integrated into most PIXE data treatment softwares like *GUPIX* [5] and [6]. This approach however, requires the depth and thickness of the intra-crystalline inclusions ($Z$ and $e$, respectively) to be measured. Several approaches have been applied so far to estimate these geometric features. One can use the optical microscope with motorized focusing to measure the depth of the top or the mid-plane of an inclusion at ±1 μm. Its floor cannot however, be clearly seen with a conventional microscope. The thickness of the inclusion is commonly taken to be equal to its width, thus assuming a spherical shape for the inclusion. Another method is confocal microscopy that allows one to precisely rebuild the 3D-geometry of an inclusion. The exact path of the beam through an inclusion during the time of PIXE data acquisition is however, very difficult to control due to a possible shift of the beam position relative to the inclusion during analysis. Several authors have proposed to deduce the depth of an intra-crystalline inclusion from the PIXE spectrum itself [7], [8] and [9]. For a given element the $K_*/K_\beta$ method consists of adjusting the parameter $Z$ so that the X-ray yields computed for the $K_*$ and $K_\beta$ lines become equal. In silicates, this method presents an acceptable accuracy only for Ca. For elements lighter than Ca, its precision is limited by the detector resolution which does not allow the $K_*$ and $K_\beta$ lines to be properly separated. For heavier elements, the difference in absorption of the $K_*$ and $K_\beta$ lines by silicates becomes insignificant. Finally Menez et al. [10] proposed to determine the $Z$ parameter from the energy at which the Na-resonance occurs in the inclusion. Beyond the fact that this method can only be applied to Na-rich inclusions, it has the drawback of significantly increasing the data acquisition time as it requires additional PIGE (particle induced $\gamma$-ray emission) measurements to be performed at different conditions from the ones required for the PIXE spectrum. This paper proposes to investigate the capability of coupled PIXE–RBS (Rutherford backscattering spectrometry) methods to analyse quantitatively intra-crystalline micro-inclusions. Several authors already underlined that much information can be obtained from accelerator-based techniques when they are applied simultaneously [11] and [12]. Conventional RBS using *-particles is a method commonly used to obtain information on depth distribution of elements in the sample surface layers up to a few μm thick. RBS is also frequently coupled with PIXE in order to normalize X-ray spectra and to analyze the major light elements of the target, undetected by PIXE. In this paper, we examine the possibility of determining depth and thickness of 10–500 μm long solid and fluid inclusions located to 5–20 μm deep in quartz from the spectrum of backscattered protons. We show that the use of RBS presents several advantages compared to other previously listed methods. Firstly it allows both the parameters $Z$ and $e$ to be simultaneously determined along the very beam path where X-ray emission is generated. Secondly it does not increase the data acquisition time.

## 2. Results and discussion

In order to test the capability of RBS to determine the geometry of micro-inclusions, we have chosen two different cases: a fluid inclusion in quartz and a monazite micro-inclusion in quartz. The main disadvantage of using protons of a few MeV to perform RBS is that the scattering cross sections in light elements are non-Rutherford. Nonetheless, by using experimental cross sections available in *SIMNRA* [13] it is possible to model the RBS spectra and determine the different geometrical parameters.

## 2.1. Experimental arrangement

Measurements have been performed in two laboratories: Laboratoire Pierre Süe (LPS) at CEA-Saclay in France [14] and at the Institute of Ion Beam Physics and Materials Research of the Forschungszentrum Dresden-Rossendorf in Germany [15]. The experimental conditions at the LPS were a proton beam of 3.2 MeV and 3.5 MeV with a beam spot of 5 μm. RBS measurements were done by an annular PIPS detector and PIXE measurements by a Ge detector. At the Rossendorf microprobe, we have used a proton beam of 3 MeV with a beam spot of 3 μm. RBS measurements were done also with a PIPS detector and PIXE with a Si(Li) detector. In both cases, beam current was of the order on the nA and measurement times were around 1 h.

## 2.2. Fluid inclusion

In the case of fluid inclusion, we have chosen a complex shape inclusion (cf. Fig 1). In this case, we see that it is necessary to determine the geometrical parameters at the very point of analysis. This inclusion has a variable width of 10–30 μm and a length of approximatively 100 μm. The depth and the thickness are variable. Fig. 1 represents the RBS spectrum of the fluid inclusion with a 3.5 MeV proton beam. The thin solid line is the spectrum that would be obtained if there was only quartz. This allows us to show the highly non-Rutherford behavior of the scattering of protons by oxygen and silicon. The effect of the fluid inclusion presence on the spectrum is the hole between 2 and 2.3 MeV and a bump between 1.7 and 2 MeV. It is due to the fact that the fluid inclusion is of course mainly composed of water. So, the hole is due to the lack of silicon and the bump to the higher amount of oxygen. By modeling this Si decreasing and this $O$ increasing, it is then relatively easy to determine the depth and the thickness of the fluid inclusion by using *SIMNRA* software. In this case, the computed depth $Z$ is of $14.4 \pm 0.5$ μm and the thickness $e$ is $11.1 \pm 0.5$ μm. The difficulty in such fluid inclusion analysis is first to have a good mechanical stability of the sample holder as one measurement can take up to 1 h as the beam current can not be too high to avoid inclusion leaking due to the heat deposited by the beam.

## 2.3. Solid inclusion

For solid inclusion we have taken the case of a micro-inclusion of monazite again in quartz (see Fig. 2 [16]. The composition of monazite is $PO_4$(Ce, La, Nd) with traces of Th, U and Pb. The interest of monazite is that it is possible to determine the age of the crystal by measuring Th, U and Pb concentrations as Pb is only coming from the disintegration of U and Th [17]. As the disintegration periods are well-known, it is easy to compute the age from the following relationship:

$$C_{Pb} = C_{Th} \frac{M_{Pb^{208}}}{M_{Th^{232}}} (e^{\lambda_{232} t} - 1) + C_{U} \frac{M_{Pb^{206}}}{M_{U^{238}}} 0.9928 (e^{\lambda_{238} t} - 1)$$
$$+ C_{U} \frac{M_{Pb^{207}}}{M_{U^{235}}} 0.0072 (e^{\lambda_{235} t} - 1)$$

where $\lambda_{232}$, $\lambda_{238}$ and $\lambda_{235}$ are, respectively, equal to $4.93 \times 10^{-11}$, $1.55 \times 10^{-10}$ and $9.85 \times 10^{-10}$ years$^{-1}$ and where $M_{Pb}^{206}$, $M_{Pb}^{207}$, $M_{Pb}^{208}$, $M_{Th}^{232}$, $M_{U}^{235}$ and $M_{U}^{238}$ are the respective isotopic masses. RBS analysis is needed to determine the inclusion geometrical parameters which are required in the analysis of the PIXE spectrum e.g. with *GUPIX* program. Fig. 2 shows the elemental mapping using RBS (b) and Th (c). Single point analysis was performed in the more intense Th region. Fig. 3 represents the 3.2 MeV proton RBS spectrum of the monazite micro-inclusion. Again the thin solid line represents the RBS spectrum of quartz alone. In this case, the monazite is on the surface and its thickness *e* is $5.9 \pm 0.5$ μm. By feeding *GUPIX* with the thickness and the matrix composition of the inclusion, Th, U and Pb concentrations were determined and are, respectively, equal to $45{,}500 \pm 500$ ppm, $5500 \pm 250$ ppm and $880 \pm 80$ ppm. The computed monazite age is then equal to $300 \pm 20$ million years. Detection of monazite micro-inclusion is easier than for a fluid inclusion due to the presence of heavy elements. However, inaccurate database of *L*-line cross section specially for Th and U can lead to high uncertainties on the age determination. A stable microbeam is again needed in this case because irradiation time is long due to the low Pb concentration.

## 3. Conclusion

We have shown in this paper that proton RBS can be used to determined the geometrical parameters of fluid and solid micro-inclusions. The optimal microprobe setup for analysis of intra-crystalline inclusions is a beam size of 1 μm, good detector efficiency and resolution to separate rare-earth lines. Special attention is to be given to the microprobe beam stability as the single point measurements are long.

## Acknowledgement

We are indebted to the Institut Universitaire des Sciences Nucléaires and to the Studium for financial supports.

## Figures

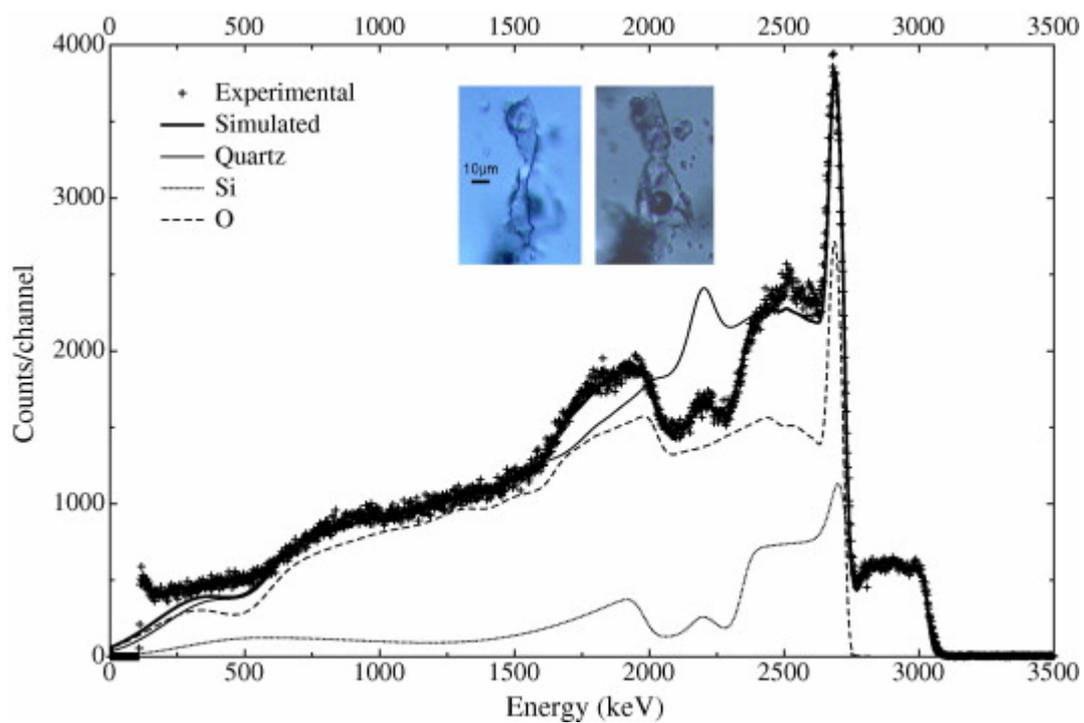

Fig. 1. Experimental and fitted 3.5 MeV proton RBS spectrum of fluid inclusion.

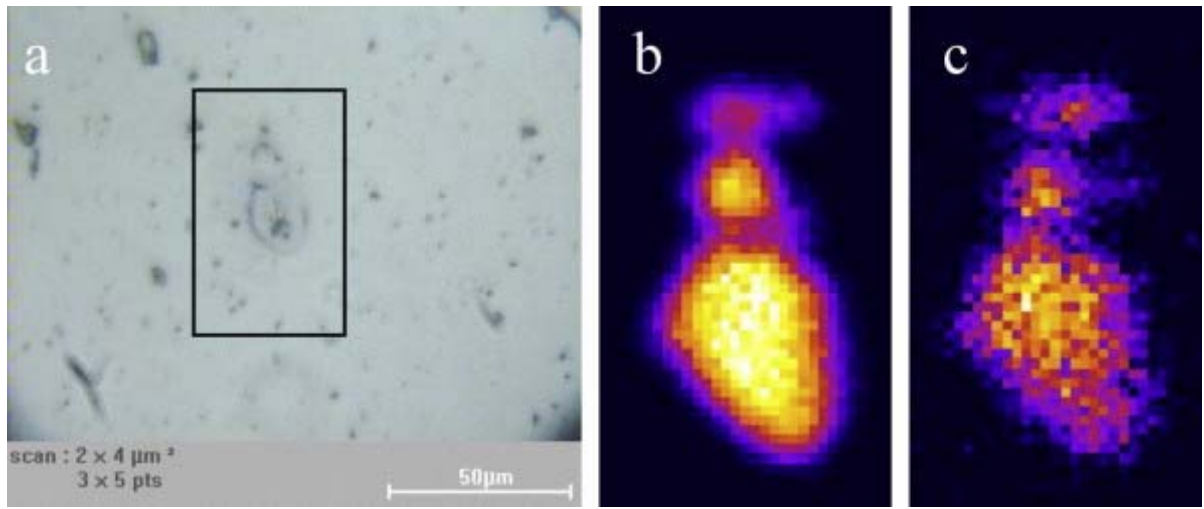

Fig. 2. Optical image (a) and elemental mapping of the monazite by the Ce, La, Nd and Th signal from RBS spectrum (b) and Th PIXE peak (c).

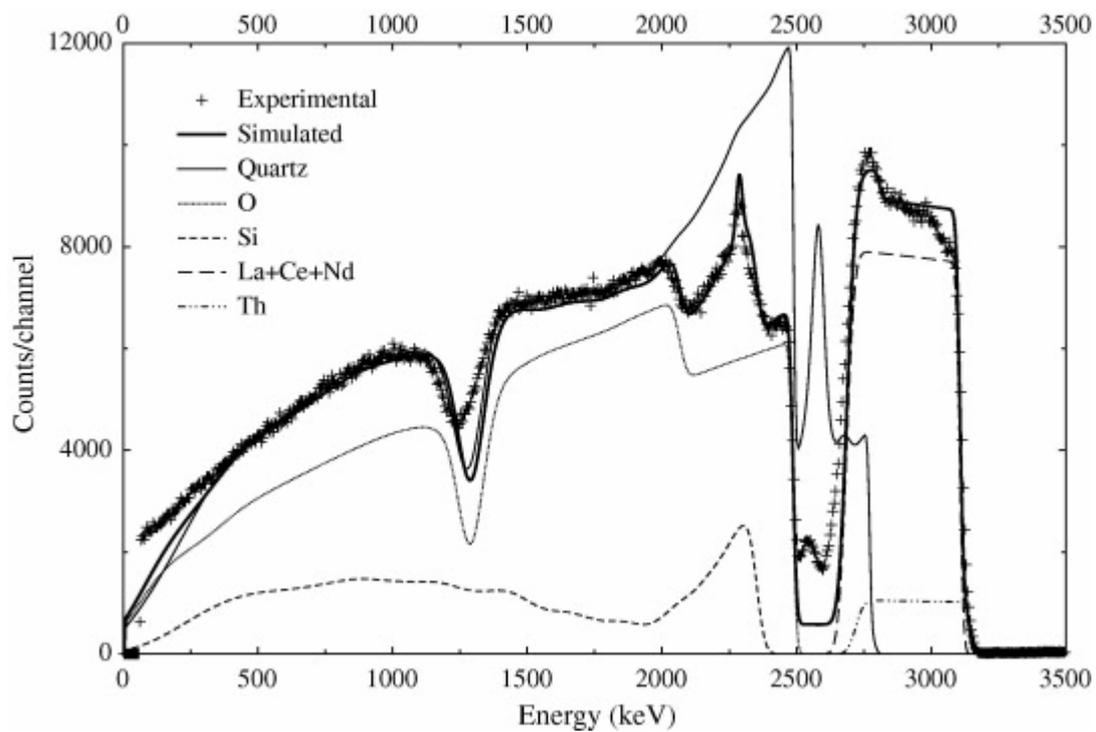

Fig. 3. Experimental and fitted 3.2 MeV proton RBS spectrum of the monazite.